\documentstyle[prl,aps,graphicx]{revtex}

\begin{document}
\draft
\date{\today}
\widetext
\title{Planck Scale Mixing and Neutrino Puzzles
\footnote{BGU PH-98/09} }
\author{David Eichler and Zakir F. Seidov
\footnote{Author to whom correspondence should
be addressed.  Electronic address: seidov@bgumail.bgu.ac.il.}
 \\
{\it {Department of Physics, Ben-Gurion University of the Negev,
Beer-Sheva 84105, Israel}}}
\maketitle 
\begin{abstract}
 It is hypothesized  that neutrinos are massless and mixed among
3 ($\nu_{\mu}$, $\nu_{e}$ and $\nu_{\tau}$) flavors with a mixing
length of order
of ${\alpha}^{-1}{l_{Planck}}{(E_{\nu}/{E_{Planck}})^{-2}}$, where
${\alpha}$ is a universal coupling constant. It is suggested that this  
hypothesis  mostly reconciles   standard solar models with
the observations of solar $^{8}B$ neutrinos, solar $pp$ neutrinos, and the
ratio of atmospheric muon neutrinos to electron neutrinos.
\end{abstract}
\pacs{PACS numbers: 14.60.Pq, 96.40.Tv}
{\em{Keywords:}} Solar neutrinos; atmospheric neutrinos; Planck
scale mixing\\
\section{Introduction}
In response to discrepancies that have persistently  been reported between
predicted and measured neutrino fluxes, some authors have suggested a
violation of Lorentz invariance, either via different
couplings of neutrinos to gravity [1], or by  different indices of
refraction of the vacuum to different types of neutrinos[2].

In this Letter we consider another such hypothesis, in 
which Lorentz invariance is violated by a foam-like vacuum 
that is granulated near the Planck scale.
 We consider that neutrinos are mixed over a scale
of ${r_{d}(E_{\nu})={\alpha}^{-1}{l_{Planck}}{(E_{\nu}/
{E_{Planck}})^{-2}}}$, where $E_\nu$
is the neutrino energy in a cosmic rest frame
and is conserved by the mixing process,
 $l_{Planck}$ $(E_{Planck})$ is the Planck
length (energy) and $\alpha$ is a dimensionless number of order
of 10$^{-2}$. We note that the hypothesis mitigates the discrepancy
between SSM predictions and the
GALLEX/SAGE, Homestake, and SuperKamiokande (SK) data for solar
neutrinos. It also mitigates the reported anomaly in  atmospheric 
neutrinos as  well. It is also marginally consistent with limits set by
post-beam dump oscillation experiments,
which reveal no significant oscillation at neutrino energies of 30
GeV over distances of order of 1/2 km.
Physical justification for the hypothesis is not quite known to us
at present, and
the results are presented here in the spirit of  suggestive 
numerical coincidence: that the energy scale in the expression for $r_d$
happens to be near the Planck scale. A brief discussion of possible
physics that could provide some theoretical justification for this
hypothesis is given in section IV.

We will, for convenience, assume  that the mixing redistributes neutrinos
uniformly and irreversibly among three flavors. Reversible mixing, that
is, oscillations, could also be
considered  at a phenomenological level with an oscillation length that
scales as $\;E^{-2}\;$, and would give similar results
(except for monochromatic neutrinos), if the coupling among all three
flavors were the same. The major feature of the present hypothesis
(although we shall refer to it for convenience as an irreversible mixing 
 [IM] hypothesis) is the energy dependence of the mixing length
$r_{d}\propto E^{-2}$, and not so much the irreversibility.

Consider a case of $n$  neutrino flavors, all interacting with 
the foam with the same mean free-path length $r_d$.
In this case the evolution of each  flavor number ${y_i}$, 
(${\Sigma}y_{i}=N=const$), as a function of
distance $\;r\;$ from the source is governed by the expression:
\begin{equation}
y_{i} = {\left(y_{i}(0)-{N \over n}\right)\exp({-r \over r_{d}})
+{N \over n}},\;\; i=1,{\;}n.
\end{equation}
Note that, if only a single flavor is emitted (as in the solar neutrino
case), then the spectrum of the emitted flavor first softens with time
(that is, with distance from source), then returns to its original form at
a lower amplitude.
\section{Solar neutrinos}
The solar neutrino problem may be described as a deficit of the
observed neutrino fluxes compared to predicted ones:\\ 
\noindent
a) The SuperKamiokanda experiment[3] presently sets the upper limit to
the Boron neutrino flux at $ 2.44{\times} 10^{-6}\; cm^{-2}\; s^{-1}$,
more than two times less than the SSM
BP98[4] "theoretical" value $5.15{\times} 10^{-6}\; cm^{-2}\; s^{-1}$;\\
\noindent
b) GALLEX and SAGE experiments[5] measure the reaction rate in $^{71}Ga$
detector as $72 \pm 5.5$ SNU (figure reported at Neutrino'98) which is
$1.8$ times less than SSM BP98[4] value
129 SNU, the Gallium  rate being caused mainly by
low-energy $pp$ neutrinos;\\
\noindent
c) The Homestake experiment[6] with $^{37}Cl$ reports a  reaction rate 
as $2.56 \pm .23$ SNU which is three 
times less than SSM BP98 value 7.56 SNU, here the detected $\nu$'s being
mostly energetic Boron neutrinos.\\
We also note that the $pp$ neutrino energy spectrum and total flux are
claimed[4]  to be known much better than those of Boron neutrinos.
\subsection{Chlorine case}
\widetext

Table I presents the results of applying the IM
hypothesis to the Solar Chlorine Case. The number of 
neutrino flavors was taken be equal to 3 with no sterile flavor. 
Individual yields in the total rate 
were calculated using
files from Web site [7]. Spectral distortion
effects are small in the Chlorine case as $pp$ neutrinos give no
yield while Boron neutrinos (due to their large energies and strong 
dependence of IM effect on the neutrino energy) give almost exactly
the limiting value
 $\;1/n=1/3\;$. That is, all $Boron$ electronic neutrinos
have been mixed into three active flavors equally.
\subsection{Gallium case}
\widetext
Table II presents the results of applying the IM hypothesis to the Solar
Gallium Case. The number of active mixing flavors of neutrinos was taken
to be equal to 3 with no sterile flavor. Spectral distortion effects are
included in the calculation of rates with the use of files in [7].
Again, as in the Chlorine case, spectral distortion effects are small as
$pp$  neutrinos have low energy (and hence large free-path length), while 
Boron neutrinos (due to their large energies and strong 
dependence of IM effect on the neutrino energy) give almost exactly
the limiting value $1/n=1/3$. 
\subsection{Seasonal variations}
IM predicts  variation in the solar
neutrino fluxes caused by the variation of the 
Earth-Sun distance . Seasonal variation of the Earth-Sun
distance, $\epsilon=\delta r/r=2e$, ($e=.017$ is the mean
eccentricity of Earth orbit), leads to relative variation of the total
$\nu_{el}$-flux, $\delta F/F =-2 \epsilon$ without IM; the IM leads
to the following relative variation of the initial flavor
$\nu$'s flux:
\begin{equation}
{\delta F \over F}=
-\epsilon \left[2+{(n-1)\over{n-1+exp{(r/r_d)}}}\right].
\end{equation}
Numerically, taking $n=3$ and $r=r_d=1 A.U.$ for $E_\nu=1 MeV$,
we get ${\delta}F/F = 8.2\% $. This value may be approximately applied
to $^{7}Be$ and $pep$ solar neutrinos
with mean energies of about $.8$ and $1.44\; MeV$, respectively. 
For $^{8}B$ neutrinos, the mean energy is
$6.7\; MeV$, and, as free-path $r_d$ is proportional to ${E_\nu}^{-2}$,
the variation of the flux is reduced to $6.8\%$. 
\section{Atmospheric neutrinos}
In this field there are two observed "deficits":\\
a) The flux of upward-going neutrinos (mainly muonic) is about .6 of flux
of downward-going neutrinos.\\ 
b) The muon-to-electron neutrino relation is observed to be about .6 
of the predicted value which is slightly larger than 2.\\
We discuss both cases.   
\subsection{Up-down asymmetry} 
Up-down asymmetry in  muonic and electronic events was observed
in a number of experiments and was suggested in [8] as a simple test for
discriminating among neutrino oscillation/mixing mechanisms and their
modifications. 
\narrowtext
\begin{figure}
\includegraphics[scale=.25]{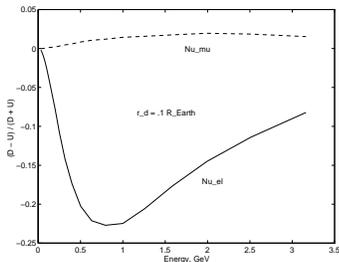}
\caption{Up-down asymmetry for atmospheric muon and electron
neutrinos, at the SK site, due to Irreversible Mixing for value
of $r_{d}=.1\;R_{Earth}$ at $E_{\nu}=1\;GeV$.}
\end{figure}
\widetext
In Fig. 1 we present the ${E_\nu}^{-2}$ mixing-induced up-down
asymmetry parameter as a function of ${E_\nu}$ for both electronic 
and muonic neutrinos, with $r_{d}=.1\;R_{Earth}$ at $E_{\nu}=1\; GeV$. 
We note that the energy dependence of the muon asymmetry and electron
asymmetry differs significantly from the cases considered in [8].     
We use atmospheric neutrino fluxes from [9] and calculate only fluxes
at the SK site
but not reaction rates for the SK detector as our aim is to show only
main characteristics of up-down asymmetry in the IM model. Note that
atmospheric
neutrino fluxes are calculated with rather large uncertainties that 
depend on the solar activity cycle phase, geomagnetic effects etc. These
uncertainties may be especialy important at low neutrino energies. 
Inclusion of these and other
factors would not drastically change the characteristics
of muonic  and electronic up-down asymmetries.    
\subsection{${\nu}_{\mu}/{\nu}_{el}$ relation}
SK experiment[10] shows a deficit of muonic events compared to
predicted values which was claimed as evidence of neutrino oscillations.
IM with values of free-path length about
$r_{d}=.1\;R_{Earth}$ or less at $E_{\nu}=1 GeV$, as it is shown in
Fig. 2, may reduce initial muonic-to-electronic-neutrino ratio down 
to observed values.
\narrowtext
\begin{figure}
\includegraphics[scale=.25]{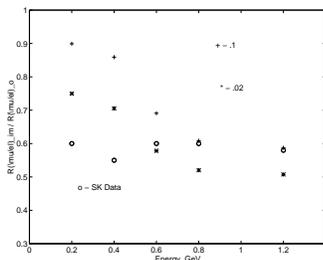}
\caption{Muon-to-electron neutrino flux ratio for atmospheric
neutrinos due to IM, at SK site, for two values
of $r_{d}$ at $E_{\nu}=1\;GeV$, .1 (+) and .02 (*). Ordinates are
ratio-of-ratios, that is, 
${\nu}_{\mu}/{\nu}_{el}$ ratio with account of IM divided by
${\nu}_{\mu}/{\nu}_{el}$ ratio without IM.
Also shown are SK[10] data on muonic-to-electronic events ratio.}
\end{figure}
\widetext
\section{Is There Any Physical Basis?}
 As  Lorentz invariance seems to be a symmetry in our low energy world,
 it is reasonable to suppose that if it is broken, such breaking
takes place at high energy, e.g. by a foam-like  condensate of quantum
black holes,
or a spaghetti-like condensate of flux tubes at the GUT scale. 
Low energy neutrino scattering off small  scale inhomogeneities so formed
would most likely have an 
scattering cross section $\propto$ E$^{2}$ , as for any
low energy neutrino scattering[11]. Also, it could be argued that
because
Lorentz invariant equations work as well as they do in describing the
observed world, with the modest price of containing infinities, then any 
violation of Lorentz invariance probably enters as an anomaly via the
choice of regularization procedure. The difficulty is that
a foam-like condensate that breaks Lorentz invariance  could also break   
translational invariance, and this would imply a scattering that is
apparently ruled out by observations of the neutrinos from 
Supernova 1987A. Also, at least for charged particles, it is ruled out by
the survival times  of particle beams in storage rings. We have no
solution to this difficulty at present; any such solution, it appears to
us, would probably require a geometric interpretation
of the different neutrino flavors as well as a model for why momentum   
and angular momentum are exactly conserved quantities whereas neutrino
flavor is not.
Alternatively, the foam may be entirely virtual, that is, unable to
increase the entropy of real particles, in which case the mixing would
be reversible. But then it is harder to justify the 
${E_\nu}^{-2}$ dependence of mixing length. 
  \section{Discussion and conclusion}
Neutrino mixing considered in this Letter 
 has only one free parameter (mean free-path length $r_d$)
given the number of flavors.
In spite of this, $E_\nu^2$ mixing
leads to a marginally acceptable account of various
deficits observed for the solar and atmospheric neutrinos.
Taking $r_{d}(E){\propto}E^{-2}$ and $r_{d}={\eta}*2\;A.U.$ at energy
$E=1\;MeV$ 
characteristic of the solar neutrino problem, then at energy $E=1\;GeV$,
characteristic of the atmospheric neutrino problem, we
have $r_{d}=({\eta}/20)\;R_{Earth}$. At $E=30\;GeV$,
characteristic of the post-beam neutrino experiments,  we have
 $r_{d}=({\eta}/3)\;km$, so the Fermilab mixing experiments[12] 
probably imply $\eta$ larger about 2.
The calculations presented in this Letter show 
that for $\eta$ about 2, the predicted count rate for the Homestake detector
is reduced from 7.5 to 3.5 SNU, the predicted rate
for GALLEX is reduced from 129 to 113 SNU, and the atmospheric 
${\nu}_{\mu}/{\nu}_{el}$ ratio is reduced by a factor of .6
at energies above .6 GeV.  Although  the reductions of the predicted 
count rate for the radiochemical detectors, Homestake, GALLEX and SAGE
 are not quite enough to completely reconcile the reported observations 
with the standard solar model, the overall trend in  these reductions for 
both atmospheric and solar neutrino experiments is perhaps suggestive 
enough to make the hypothesis   
worthy of consideration at present.

A clean prediction, however, is that longer terrestrial lengths for post
beam dump neutrino experiments must reveal mixing. A null result
over the CERN-MACRO distance or the Fermilab-Wisconsin distance would
unequivocally rule out the hypothesis. 
\section{Acknowledgements}
DE acknowldeges the IBS for partial support  and ZS is grateful to
BGU and particularly to DE for the generous hospitality.
 
\narrowtext
\begin{table}
\caption{Irreversible Mixing, with SSM BP98[4] predictions for Chlorine
reaction rates, at n=3 flavors for some values of the free-path
length $r_{d}\;$ at $\;E_{\nu}=1\;MeV$. 
Case $r_{d}=\infty\;$ corresponds to SSM BP98 predictions.}
\begin{tabular}{cccccc}
  & Source & & & Rates (SNU)&\\
  &        & ---&---&---&---\\
  &        & & & $r_d\;$(A.U.)&\\
  &   & $\infty$ &5 &4 &3\\
\tableline
1 & $pp$ & 0 & 0 & 0 & 0 \\
2 & $pep$ & .22 & .17 & .16&.15\\
3 & $^{7}Be$ & 1.15 & 1.05 & 1.03 & 1.0\\
4 & $^{8}B$ & 5.72 & 1.91 & 1.91 & 1.91\\
5 & $^{13}N$ & .10 & .09 & .09 & .08 \\
6 & $^{15}O$ & .36 & .32 & .31 & .29\\
\hline
&Total & 7.56&3.54&3.50& 3.43 
\end{tabular}
\end{table}
\narrowtext
\begin{table}
\caption{Irreversible Mixing, with SSM BP98 predictions, for Gallium
reaction rates at n=3 flavors for some values of the free-path
length $r_{d}\;$ at $\;E_{\nu}=1\;MeV$.Case $r_{d}=\infty\;$ corresponds
to SSM BP98 predictions.}
\begin{tabular}{cccccc}
  & Source & & & Rates (SNU)&\\
  &        & ---&---&---&---\\
  &        & & & $r_d\;$(A.U.)&\\
  &   & $\infty$ & 4 & 3 & 2\\
\tableline
1 & $pp$ & 69.6 & 67.6 & 68.0 & 67.1 \\
2 & $pep$ & 2.8 & 2.1 & 1.9 & 1.6\\
3 & $^{7}Be$ & 34.4 & 30.9 & 29.9  & 27.9\\
4 & $^{8}B$ & 12.4 & 4.1 & 4.1 & 4.1\\
5 & $^{13}N$ & 3.7 & 3.4 & 3.3 & 3.1 \\
6 & $^{15}O$ & 6.0 & 5.2 & 4.9 & 4.5\\
\hline
&Total & 129.0 & 113.3 & 112.1 & 108.3 
\end{tabular}
\end{table}
\end{document}